\documentclass{aa}  
\usepackage{natbib}
\bibpunct{(}{)}{;}{a}{}{,} 
\usepackage{graphicx}
\usepackage{txfonts}

\begin{document}
\title{Effects of CO$_2$ on H$_2$O  band profiles and band strengths in mixed H$_2$O:CO$_2$ ices}
\titlerunning{Effects of CO$_2$ on H$_2$O  band profiles and band strengths}

\author{Karin I. \"Oberg \inst{1}
          \and Helen J. Fraser \inst{2}
          \and A. C. Adwin Boogert \inst{3}
          \and Suzanne E. Bisschop  \inst{1}
          \and Guido W. Fuchs \inst{1}
          \and Ewine F. van Dishoeck  \inst{4}
          \and Harold Linnartz  \inst{1}
          }

   \offprints{Karin I. \"Oberg, oberg@strw.leidenuniv.nl }
 
  \institute{Sackler Laboratory for Astrophysics, Leiden Observatory, University of Leiden, P.O. Box 9513, NL 2300 RA Leiden, The Netherlands \and Department of Physics, Scottish Universities Physics Alliance (SUPA), University of Strathclyde, John Anderson Building, 107 Rottenrow East, Glasgow G4 ONG, Scotland \and AURA/NOAO-South, Casilla 603, La Serena, Chile, S. A. \and Leiden Observatory, University of Leiden, P.O. Box 9513, NL 2300 RA Leiden, The Netherlands}

   \date{Received ; accepted}

 
  \abstract
   {H$_2$O is the most abundant component of astrophysical ices. In most lines of sight it is not possible to fit both the H$_2$O 3 $\mu$m stretching, the 6 $\mu$m bending and the 13 $\mu$m libration band intensities with a single pure H$_2$O spectrum. Recent Spitzer observations have revealed CO$_2$ ice in high abundances and it has been suggested that CO$_2$ mixed into H$_2$O ice can affect the positions, shapes and relative strengths of the 3 $\mu$m and 6 $\mu$m bands. }
   {We investigate whether the discrepancy in intensity between H$_2$O bands in interstellar clouds and star forming regions can be explained by CO$_2$ mixed into the observed H$_2$O ice affecting the bands differently.}
   {Laboratory infrared transmission spectroscopy is used to record spectra of H$_2$O:CO$_2$ ice mixtures at astrophysically relevant temperatures and composition ratios.}
   {The H$_2$O peak profiles and band strengths are significantly different in H$_2$O:CO$_2$ ice mixtures compared to pure H$_2$O ice. The ratio between the strengths of the 3 $\mu$m and 6 $\mu$m bands drops linearly with CO$_2$ concentration such that it is 50\% lower in a 1:1 mixture compared to pure H$_2$O ice. In all H$_2$O:CO$_2$ mixtures, a strong free-OH stretching band appears around 2.73 $\mu$m, which can be used to put an upper limit on the CO$_2$ concentration in the H$_2$O ice. The H$_2$O bending mode profile also changes drastically with CO$_2$ concentration; the broad pure H$_2$O band gives way to two narrow bands as the CO$_2$ concentration is increased. This makes it crucial to constrain the environment of H$_2$O ice to enable correct assignments of other species contributing to the  interstellar 6 $\mu$m absorption band. The amount of CO$_2$ present in the H$_2$O ice of B5:IRS1 is estimated by simultaneously comparing the H$_2$O stretching and bending regions and the CO$_2$ bending mode to laboratory spectra of H$_2$O, CO$_2$, H$_2$O:CO$_2$ and HCOOH.}
   {}

   \keywords{Astrochemistry, Line: profiles, Molecular data, Molecular processes, Methods: laboratory, ISM: molecules, Infrared: ISM}

   \maketitle
%

\section{Introduction}

Infrared spectroscopy towards dense molecular clouds and young stellar objects (YSOs) often reveals prominent bands attributed to H$_2$O ice. The 3.08 $\mu$m (3250 cm$^{-1}$) band, attributed to the H$_2$O stretching mode, and the 6.0 $\mu$m (1700 cm$^{-1}$) band, attributed to the H$_2$O bending mode, are detected in many lines of sight \citep{willner82,tanaka90,murakawa00,boogert00,keane01,Gibb04,knez05}. Also the observation of the H$_2$O libration mode at 13 $\mu$m (770 cm$^{-1}$) has been reported \citep{keane01,Gibb04}. It is a long-standing problem in the interstellar community that the H$_2$O ice 6.0/3.08 band intensity ratio in astrophysical observations is up to 2 times higher than expected for pure H$_2$O ice \citep{keane01}. 

This discrepancy has previously been explained by absorptions of other volatile molecules and organic refractory material absorbing around 6 $\mu$m \citep{schutte96,keane01,Gibb02}. These theories were put forward to explain the strong absorption of the 6.0 $\mu$m feature as well as why its shape does not match that of pure H$_2$O ice very well in many lines of sight. \citet{keane01} identified two additional components in the H$_2$O 6.0 $\mu$m bending mode region, centered around 5.83 $\mu$m (1720 cm$^{-1}$) and 6.2 $\mu$m (1600 cm$^{-1}$), when subtracting a pure H$_2$O ice spectrum that was fitted to the 3.08 $\mu$m stretching band. Analyzing the residuals after subtracting the H$_2$O bending mode presupposes that the profile of the H$_2$O bending mode is well known. Hence knowledge of the H$_2$O bending mode profile in different environments is critical to correctly assign other species contributing to the 6 $\mu$m band.

\citet{Pontoppidan05} showed that for observations of ices in circumstellar disks, part of the discrepancy in intensities between the H$_2$O bands may be due to disk geometry because of scattering at short wavelengths. However, since this anomaly is almost always present to some degree in disks and envelopes, as well as in clouds, it is unlikely to be the entire explanation. In the sources where the 13 $\mu$m H$_2$O libration band is visible, it is possible to fit the H$_2$O stretching and libration peaks with a single H$_2$O ice abundance, while fitting the 6 $\mu$m band independently results in a much higher column density \citep{Gibb04}. The fact that the libration mode at 13 $\mu$m is in agreement with the 3 $\mu$m band refutes the idea that the excess at 6 $\mu$m is due to wavelength dependent scattering. 

Recently \citet{knez05} suggested that the ratio in band intensity could be due to large amounts of CO$_2$ mixed in with the H$_2$O ice. Observations reveal that solid CO$_2$ is common in many lines of sight \citep{gerakines99,Gibb04}. With the Infrared Space Observatory (ISO), the CO$_2$ stretching mode at 4.25 $\mu$m (2350 cm$^{-1}$) was observed toward Taurus background stars \citep{whittet98, nummelin01}. More recently the launch of the Spitzer Space Telescope made the CO$_2$ bending mode at 15 $\mu$m (670 cm$^{-1}$) available for observations and the band has been detected towards several background stars \citep{bergin05,knez05}. The average abundance with respect to H$_2$O ice towards the Taurus sources is 20\%, but up to 37\% has been observed (Knez et al., in prep.). Toward several protostars up to 35\% of CO$_2$ compared to H$_2$O has been observed \citep{nummelin01,boogert04}, making CO$_2$ one of the most abundant ices after H$_2$O.

In a previous study, using the H$_2$O column density from the 3.08 $\mu$m band and laboratory spectra of pure H$_2$O ice, \citet{knez05} determined that the H$_2$O bending mode contributes 77\% and 69\% to the observed 6.0 $\mu$m absorption features towards Elias 16 and CK 2, respectively. Using a combination of laboratory spectra of two H$_2$O:CO$_2$ mixtures, 1:1 and 10:1 respectively, they showed that 85\% to 100\% of the 6.0 $\mu$m band strength can be explained by H$_2$O. This is due to the smaller ratio between the stretching and bending mode strengths seen in an unpublished H$_2$O:CO$_2$ 1:1 spectrum by Ehrenfreund (private communication). This combination of spectra, together with a water-poor mixture spectrum, also fits the CO$_2$ profile well, with approximately 80\% of the CO$_2$ in the water rich ice \citep{knez05}. Hence it is not unlikely that a significant part of the H$_2$O ice is in H$_2$O:CO$_2$ ice mixtures close to 1:1 in many lines of sight, even if the average abundances of H$_2$O and CO$_2$ is closer to 3:1.

The question that prompted this study is whether there is a change in H$_2$O band profiles and relative H$_2$O band strengths in ice mixtures compared to pure H$_2$O ice, for H$_2$O:CO$_2$ ice mixtures that are both astrophysically relevant and contain enough H$_2$O to observe with e.g. Spitzer. Such information in turn, is a prerequisite for determining whether additional species contribute to the 6 $\mu$m band. In this work we present a systematic study of the infrared properties of H$_2$O absorptions in H$_2$O:CO$_2$ ice mixtures around 1:1, for the temperature range of 15 to 135 K, in order to constrain the effect that CO$_2$ has on the shapes and relative band strengths of the H$_2$O bands. 

\section{Previous laboratory data}

Two previous studies have reported changes in the H$_2$O bands in H$_2$O:CO$_2$ ice mixtures compared to pure H$_2$O ices \citep{hagen83,schmitt89}. In both cases, only one isolated H$_2$O:CO$_2$ mixture, 1:2 and 10:1 respectively, was investigated. In both the 1:2 and 10:1 H$_2$O:CO$_2$ mixture spectra a new H$_2$O band appears around 2.74 $\mu$m, the relative strength of the bending mode increases and all the band profiles change compared to a pure H$_2$O ice spectrum. No attempts were made to quantify these changes.

A number of later laboratory studies focused on H$_2$O:CO$_2$ ice mixtures as well, but to our knowledge none of them systematically studied the impact of CO$_2$ on water ice and none of them has reported on changes in the H$_2$O band profiles and band strengths due to CO$_2$. This is not surprising, since most of the studies focused on mixtures with H$_2$O as the dominant ice component \citep{hudgins93, bernstein05}. Only a few laboratory spectra of H$_2$O:CO$_2$ mixtures close to 1:1 exist in the literature \citep{gerakines95, palumbo00, gerakines05}. The effect of CO$_2$ on H$_2$O ice features is only considered by \citet{gerakines05}, who concluded from the spectra of a H$_2$O:CO$_2$ 1.6:1 ice mixture that the relative H$_2$O band intensities were not affected by high concentrations of CO$_2$. Nevertheless, the 3.08/6.0 $\mu$m band strength ratio can be calculated from their reported integrated intensities and reveals a drop of $\sim$30\% compared to pure H$_2$O.

It is well known from matrix isolation experiments that the different H$_2$O bands have different relative band strengths dependent on H$_2$O cluster size \citep{vanthiel57}. The band strength of the bending mode is much less affected by the H$_2$O cluster size than the stretching mode, when the H$_2$O cluster size is changed by varying the H$_2$O concentration in a N$_2$ matrix. The former band strength drops by ~20\% when the matrix to H$_2$O ratio is decreased by one order of magnitude while the stretching band strength increases by a factor of 10. The main intensity contribution in both bands comes from the monomer peak when the H$_2$O concentration is low and from a cluster mixture when the concentration is high (high concentration meaning N$_2$:H$_2$O 1:10, \citealt{vanthiel57}). The position of the bending modes is approximately the same for all cluster sizes, while the major stretching mode peak shifts from 3400 cm$^{-1}$ at high H$_2$O concentrations to the position of the free-OH stretch at low concentrations ($\sim$3690 cm$^{-1}$ or $\sim$2.7 $\mu$m). Finally, the relative band strengths are also affected by the type of matrix used, e.g. noble gas or nitrogen or oxygen. This has been reported also for H$_2$O in astronomically relevant matrices \citep{ehrenfreund96}. It is important to note that none of these matrices forms hydrogen bonds with H$_2$O, thus no conclusions about the band strengths and band profiles of H$_2$O can be drawn from these experiments concerning H$_2$O in a hydrogen bonding matrix. Another astrophysically relevant molecule, CO, does probably form hydrogen bonds with H$_2$O in amorphous ice mixtures \citep{schmitt89}. Matrix experiments have also shown that CO forms weak hydrogen bonds, while CO$_2$ does not \citep{tso85}.

\section{Experiment and data analysis}

\begin{table}
\begin{minipage}[t]{\columnwidth}
\caption{Ice mixtures studied in this work}
\label{table1}
\centering
\renewcommand{\footnoterule}{}
\begin{tabular}{l c c c}
\hline\hline
 Composition&H$_2$O (L)\footnote{1 L (Langmuir) = 1 $\times$ 10$^{-6}$ Torr s $\approx$ 1 monolayer of molecules}&CO$_2$ (L)&Total ice exposure (L)\\
\hline
pure H$_2$O&3000&0&3000\\
pure CO$_2$&0&3000&3000\\
H$_2$O:CO$_2$ 1:0.25&3000&750&3750\\
H$_2$O:CO$_2$ 1:0.5&3000&1500&4500\\
H$_2$O:CO$_2$ 1:1&3000&3000&6000\\
H$_2$O:CO$_2$ 1:2&3000&6000&9000\\
H$_2$O:CO$_2$ 1:4&3000&12000&15000\\
H$_2$O:CO$_2$ 1:1&10000&10000&20000\\
H$_2$O:CO$_2$ 1:1&1000&1000&2000\\
\hline
\end{tabular}
\end{minipage}
\end{table}

\begin{table}
\begin{minipage}[t]{\columnwidth}
\caption{The measured peak positions and the integration bounds in cm$^{-1}$ ($\mu$m) used to compute the integrated intensities of the H$_2$O peaks}
\label{table2}
\centering
\renewcommand{\footnoterule}{}
\begin{tabular}{l c c c}
\hline\hline
 &&\multicolumn{2}{c}{Integration bounds}\\
H$_2$O bands&Peak&Lower&Upper\\
\hline
libration&780 (12.8)&500 (20.0)&1100 (9.09)\footnote{The CO$_2$ bending mode is excluded by explicitly taking into account only half of the libration mode profile (see text)}\\
bend&1655 (6.04)&1100 (9.09)&1900 (5.26)\\
bulk stretch&3279 (3.05)&3000 (3.33)&3600 (2.78)\\
free OH stretch&3661 (2.73)&3600 (2.78)&3730 (2.68)\\
\hline
\end{tabular}
\end{minipage}
\end{table}

\subsection{Experiment}

All experiments were conducted in a high vacuum (HV) chamber described in detail elsewhere \citep{gerakines95} at a base pressures below 1.3$\times$10$^{-6}$ Torr at room temperature. Ices of C$^{18}$O$_2$ (Praxair 97\% purity) and H$_2$O (deionized and further purified through sequential freeze-thawing in a vacuum manifold) were grown on a CsI window, precooled to 15 K (45 K for one specific experiment), via effusive dosing at a growth rate of $\sim$10$^{16}$ molecules cm$^{-2}$ s$^{-1}$ roughly along the surface normal. C$^{18}$O$_2$ was used instead of the main isotopologue of  CO$_2$ to minimize overlap between H$_2$O and CO$_2$ spectral features. Transmission Fourier transform infrared spectra of the ice systems were recorded between 4000 - 400 cm$^{-1}$ (2.5 - 25 $\mu$m) at a spectral resolution of 2 cm$^{-1}$ at fixed temperatures between 15 and 135 K, using a total of 256 scans. Background spectra were acquired prior to deposition for each experiment, at the same resolution and number of scans, and automatically subtracted from the recorded ice spectra. 

The pure ices were grown {\it in situ} from C$^{18}$O$_2$ and H$_2$O gas bulbs that were filled to  a total pressure of 10 mbar, prepared in a glass-vacuum manifold at a base-pressure of $\sim$10$^{-4}$ mbar. Mixed ices were made by dosing gas from premixed H$_2$O:C$^{18}$O$_2$ bulbs, also prepared in the glass-vacuum manifold. 

The ice mixtures studied here are summarized in Table \ref{table1}. The relative concentrations for the H$_2$O:C$^{18}$O$_2$ mixtures range from 20 to 80\% CO$_2$, where the \% are relative to the whole ice, i.e. 20\% CO$_2$ is equivalent to a H$_2$O:CO$_2$ 4:1 mixture. The H$_2$O ice exposure was always kept at 3000 L (1 L (Langmuir) = 1 $\times$ 10$^{-6}$ Torr s $\approx$ 1 monolayer of molecules, assuming 10$^{15}$ molecules cm$^2$ and a sticking probability of 1), except in two 1:1 control experiments with 1000 and 10000 L H$_2$O exposure, respectively. The nomenclature adopted is as follows, A:B denotes a mixture with A parts of H$_2$O and B parts of C$^{18}$O$_2$. 

\subsection{Data analysis}

\begin{figure}
\resizebox{\hsize}{!}{\includegraphics{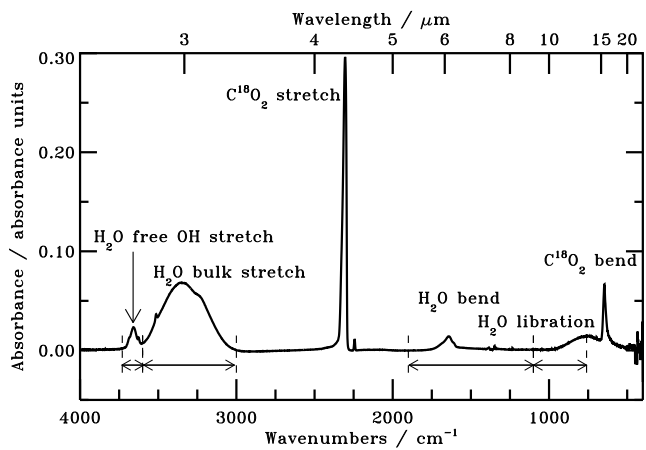}}
\caption{Spectra over the 4000-400 cm$^{-1}$ range of the H$_2$O:CO$_2$ 1:0.5 ice mixture at 15 K. The assignments of the major peaks are indicated.  The double headed arrows indicate the integration ranges for determining the intensities of the various bands; for the libration mode, only half the range is indicated (see text). Note that all CO$_2$ frequencies are shifted compared to those observed in space, since in the present experiments C$^{18}$O$_2$ has been used.}
\label{fig1}
\end{figure}

The acquired spectra were first reduced to flatten the baseline by fitting a second order polynomial to the same 5 points, chosen by visual inspection well away from any features. This was done to facilitate comparison between laboratory and astronomical spectra. The curved baseline in the raw data is a real feature which is due to scattering within the ice. Flattening the spectra may hence distort profiles, but in our case no such distortion was noted. All reduced spectra can be found at http://www.strw.leidenuniv.nl/$\sim$lab/databases.

Figure \ref{fig1} shows an overview spectrum of a H$_2$O:CO$_2$ 1:0.5 mixture. The use of the C$^{18}$O$_2$ isotope changes the position of the CO$_2$ bands compared to the main isotopologue, but does not affect the conclusions of this study. In addition to the C$^{18}$O$_2$ bending and stretching modes, there are two weak, narrow C$^{18}$O$_2$ overtone bands in the spectrum, at 3515 and 3625 cm$^{-1}$ respectively, that overlap with the H$_2$O stretching bands. To calculate the integrated intensities of the H$_2$O bands, the H$_2$O peak intensities were integrated over the same wavenumber range for all spectra (see Table \ref{table2} and Fig. \ref{fig1}). Where narrow CO$_2$ peaks overlapped with the H$_2$O bands, a Gaussian was fitted to the CO$_2$ peak and subtracted from the spectrum. A Gaussian could not be fitted to the CO$_2$ bending mode, which overlaps with the H$_2$O libration mode. The band strength of the H$_2$O libration band was instead calculated by doubling the band strength of the high frequency half of the band. The peak frequency of the H$_2$O libration mode varies in the different mixtures and since the lower frequency integration bound was set to the peak frequency, the lower bound is somewhat different for different mixtures. For pure H$_2$O ice, this procedure was found to be accurate to within 3\% compared to integrating the whole profile. To further test the induced error of this approach the band strengths were also calculated by simply subtracting the integrated area of the CO$_2$ bending mode from the total integrated area. The difference was less than 5\%. The large interval required for the H$_2$O bending mode is due to its substantial low frequency wing attributed to the librational overtone \citep{devlin01}, which in some spectra contains more than 50\% of the integrated peak intensity. The 2.73 $\mu$m band is due to free-OH stretches and its assignment to H$_2$O monomers, dimers and small multimers is justified in the discussion part of this paper.

The substantial change in the H$_2$O band intensities between different mixtures reported here, prevents an independent check of the relative amount of H$_2$O deposited onto the surface. From three repeated experiments (of the 2:1 mixture) the standard deviations of the integrated H$_2$O peak intensities are estimated to be less than 10\%. These experiments were carried out in the beginning and at the end of an experimental series and for separately prepared mixtures. The standard deviation hence contains the error from mixing, absolute ice exposure, measuring errors and changes in the experiment over time. Additional errors may arise from the flattening of the spectra; the difference between integrated intensities for the raw and reduced spectra is the largest for the 6.0 $\mu$m H$_2$O band at low H$_2$O concentrations; up to 40\% for the 1:4 mixture. For the astrophysically relevant mixtures this uncertainty is only 1 - 5\%, however. Some systematic errors due to the mixing procedure cannot be excluded, but are difficult to quantify. Taking all error sources into account, we estimate that the relative band strengths are accurate within $\sim$10\% for the astrophysically relevant ice mixtures, i.e. ice mixtures with less or equal amounts of CO$_2$ compared to H$_2$O. 

The band strengths, $A$, of the three H$_2$O bands present in pure H$_2$O ice were estimated for all mixtures using the measured band strengths for pure H$_2$O ice at 14 K by Gerakines et al. (1995). The band strengths for the pure H$_2$O bands were thus set to 2.0 $\times$ 10$^{-16}$ for the H$_2$O stretching mode, 1.2 $\times$ 10$^{-17}$ for the bending mode and 3.1 $\times$ 10$^{-17}$ cm molecule$^{-1}$ for the libration band. The band strengths of these bands in the H$_2$O:CO$_2$ mixtures were calculated by scaling each integrated intensity by the band strength of the pure H$_2$O ice band over the pure H$_2$O band integrated intensity:

 \begin{equation}
A_{\rm H{_2}O:CO{_2}{=}1:x}^{\rm band}{=}{{\int_{band}{I_{\rm H{_2}O:CO{_2}{=}1:x}}}{\times} \frac{ A_{\rm H{_2}O}^{\rm band}}{\int_{\rm band}{I_{\rm H{_2}O}}}} 
 \label{eqn1}
 \end{equation}

\noindent where ${A_{\rm H{_2}O:CO{_2}{=}1:x}^{\rm band}}$ is the calculated strength of each H$_2$O band in the 1:x H$_2$O:CO$_2$ mixture, $\int_{\rm band}{I_{\rm H{_2}O:CO{_2}{=}1:x}}$ the measured integrated intensity of the same band, ${A_{\rm H{_2}O}^{\rm band}}$ the known strength of the pure H$_2$O band and $\int_{\rm band}{I_{\rm H{_2}O}}$ the integrated intensity of the pure H$_2$O band. The band strengths of the free OH stretch in the different mixtures were scaled to the band strength of the bulk stretching mode in pure H$_2$O ice. The ratios of our measured integrated intensities for pure H$_2$O at 15 K coincided with those of Gerakines et al. within 10\%. 

The calculated strengths of all H$_2$O bands were plotted as a function of CO$_2$ concentration in the ice mixture and fitted by linear models as described in detail in section 4.1. These models are more accurate in predicting the band strengths for H$_2$O in a certain mixture than individual measurements, since they are derived from all experiments and hence the random errors are averaged out. Our model predictions agreed well with previously published isolated measurements. A spectrum by Schutte (Leiden Molecular Database) of H$_2$O:CO$_2$ 1:1.25 has a stretching to bending peak ratio which lies within 5\% of the value predicted by our model fit. In addition, the ratio of stretching to bending modes in the H$_2$O:CO$_2$ 1.6:1 mixture from \citet{gerakines05} lies within 8\%  of our model value, further corroborating the results presented in this paper.

\section{Results}

\subsection{Changes in H$_2$O band strengths and profiles with mixture composition}

\begin{figure*}
\centering
\includegraphics[width=17cm]{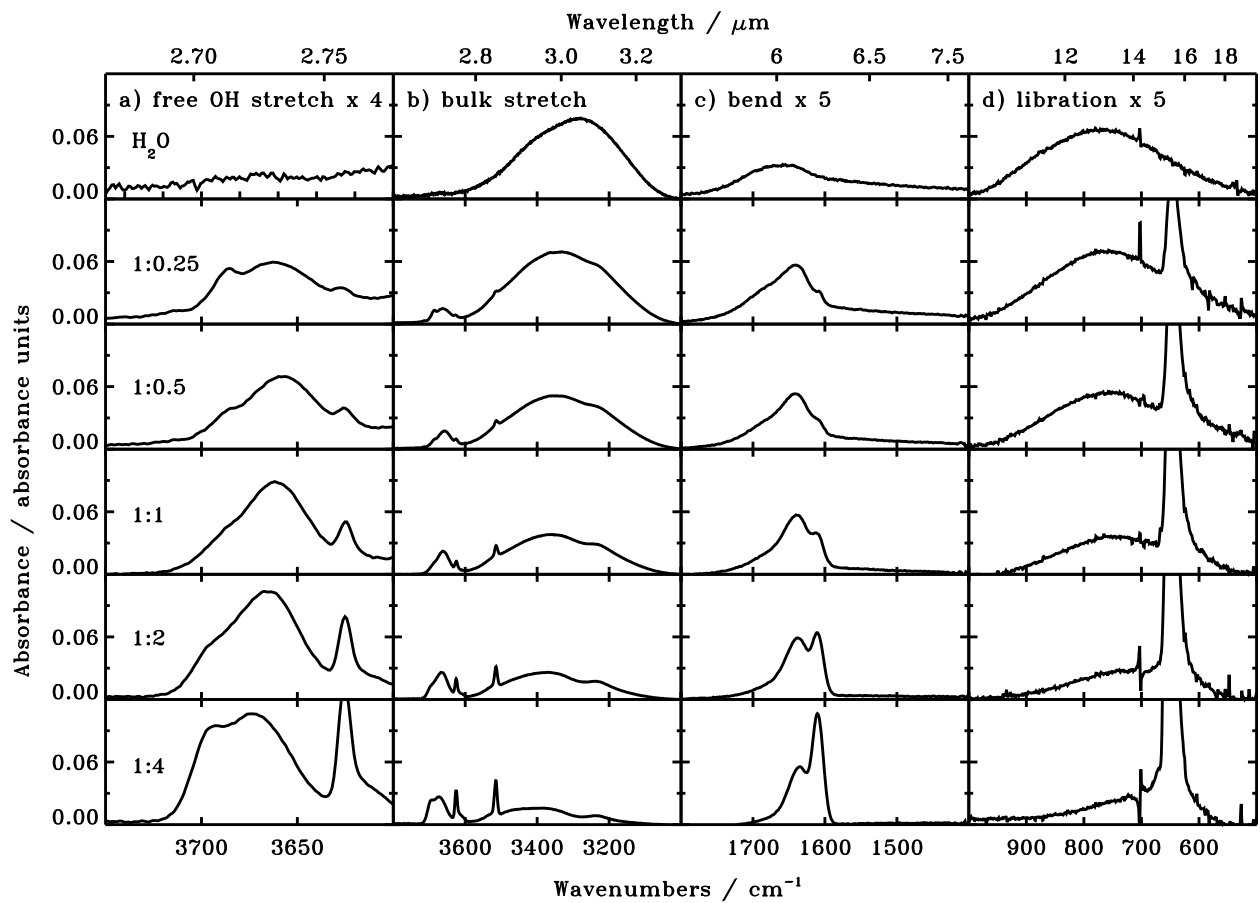}
\caption{Spectra over the 4000-400 cm$^{-1}$ range at 15 K of the four H$_2$O peaks in H$_2$O:CO$_2$ 1:x ice mixtures for different CO$_2$ concentrations. Each column contains one of the H$_2$O modes and each row one of the H$_2$O:CO$_2$ mixtures: a) an expanded view of the H$_2$O free OH stretch, b) the H$_2$O bulk and free-OH stretch, c) the H$_2$O bend and  d) H$_2$O libration. The intensities of the libration, bend and free-OH stretch bands have been scaled, with a scaling factor indicated in the first row. The H$_2$O exposure was kept constant in all experiments. The narrow peaks around 3500 and 3630 cm$^{-1}$ have been previously assigned to CO$_2$ combination modes \citep{sandford90}. The sharp feature around 700 cm$^{-1}$ is an experimental artifact.}
\label{fig2}
\end{figure*}

Figure \ref{fig2} shows that the H$_2$O spectra at 15 K undergo two significant changes as the amount of CO$_2$ is increased from 0 to 80\%; the profiles of the H$_2$O bending and stretching bands change dramatically and all H$_2$O integrated peak intensities change systematically with varying CO$_2$ concentration. 

\begin{figure}
\resizebox{\hsize}{!}{\includegraphics{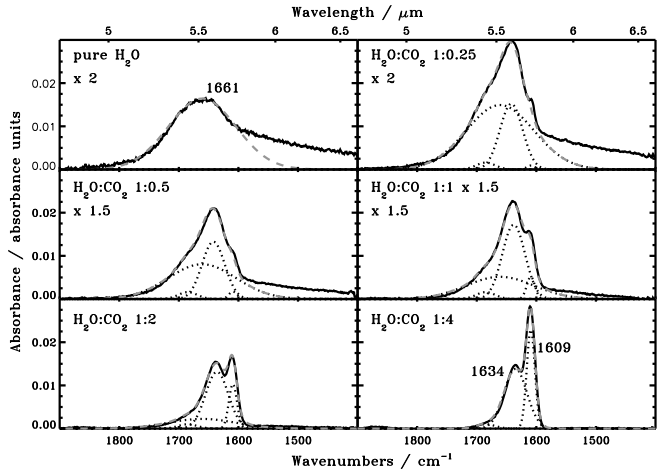}}
\caption{The three Gaussian components present in the H$_2$O bending mode in mixtures with CO$_2$ disregarding the low frequency wing. In the upper row, first column the pure H$_2$O bending mode consists of a wing and a peak, where the peak was fitted with a Gaussian at 1661 cm$^{-1}$. In the lower row, second column, the H$_2$O bending mode in the 1:4 mixture can be separated into two narrow Gaussians at 1609 and 1634 cm$^{-1}$. The four other H$_2$O:CO$_2$ mixtures can be separated into these three components derived from the pure and 1:4 mixed ice. A small additional Gaussian centered at 1685 cm$^{-1}$ accounts for some high frequency excess. The dotted lines indicate the individual Gaussian components and the dashed lines their sum.}
\label{fig3}
\end{figure}

\begin{figure}
\resizebox{\hsize}{!}{\includegraphics{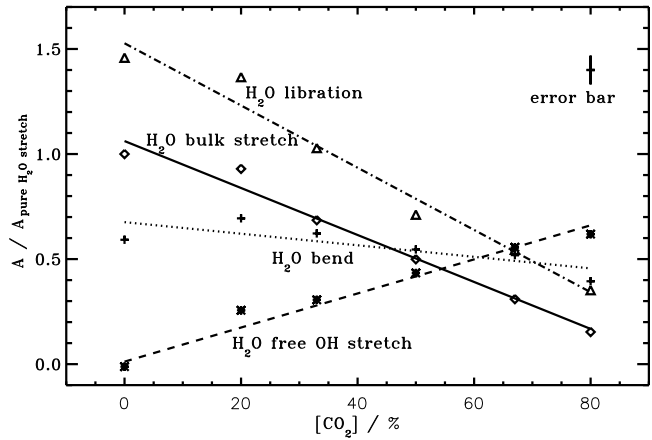}}
\caption{The band strengths of the four main H$_2$O peaks in H$_2$O:CO$_2$ ice mixtures relative to that of the H$_2$O stretch in pure H$_2$O ice at 15 K. The band strengths of the libration, bend and free OH stretch modes have been multiplied by a factor of 10 to facilitate display. The estimated average error bar of the relative band strengths is shown in the upper right corner.}
\label{fig4}
\end{figure}

\begin{figure}
\resizebox{\hsize}{!}{\includegraphics{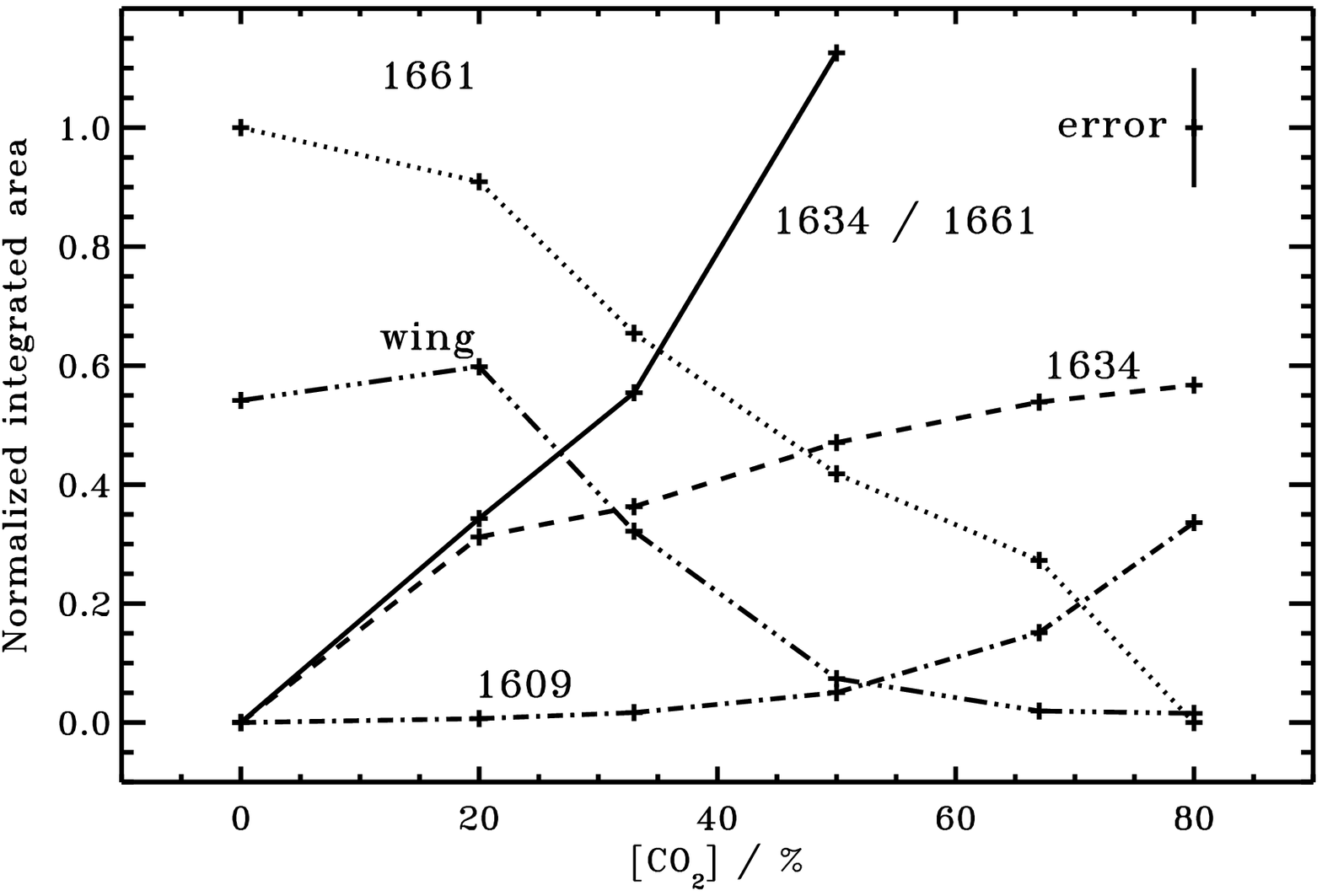}}
\caption{The integrated area of the three Gaussian components present in the H$_2$O bending mode in mixtures with CO$_2$ and of the low frequency wing attributed to a librational overtone. The ratio between the 1634 and 1661  cm$^{-1}$ peaks is plotted as well. The estimated error, shown as an error bar in the top right corner, is mainly due to the error in the total band strength and not to the Gaussian fit.}
\label{fig5}
\end{figure}

Of all modes, the profile the H$_2$O bending mode is most affected by the CO$_2$. In pure H$_2$O the bending mode consists of a broad peak centered at 1661 cm$^{-1}$ (6.02 $\mu$m). The bending mode overlaps with the librational overtone, which shows up in the spectra as a low-frequency wing. To simplify comparison with astronomical observations, the wing is generally treated as a part of the bending mode when calculating band strengths. The band strength of the pure bending mode is smaller for the mixtures that have strong librational modes and hence strong librational overtones. In the spectrum of the 1:4 mixture, two narrow peaks are observed in the H$_2$O bending region at 1609 and 1634 cm$^{-1}$, instead of the broad peak and wing in the pure H$_2$O spectrum. The profiles of the bending mode in the other mixtures appear to be a composition of the peaks in the pure H$_2$O spectra and the H$_2$O:CO$_2$ 1:4 mixture. To test this, Gaussian profiles were fitted to the bending peak of the pure H$_2$O ice, disregarding the wing,  and to the two peaks in the 1:4 mixture (Lorentzian profiles were also attempted but were impossible to match to the ice peaks). The IDL simplex optimization routine was used to fit two Gaussian peaks simultaneously to the band in the 1:4 spectrum. Figure \ref{fig3} shows that the bending mode in all other mixtures can be separated into these three peaks; the slight misfit on the blue wing in all mixtures is corrected for by an additional, small Gaussian centered at 1685 cm$^{-1}$ with FWHM (Full Width at Half Maximum) 35 cm$^{-1}$. The position of the 1634 cm$^{-1}$ peak is shifted with CO$_2$ concentration (2 to 8 cm$^{-1}$ for the different mixtures), but the FWHMs of the Gaussians were always kept constant. The other peak positions were also kept constant. The positions, widths and intensities of the Gaussians used to fit the bending mode for each mixture are listed in Table \ref{table3}. 

The same bulk stretching peak apparent in the pure H$_2$O spectra is visible in all mixture spectra (Fig. \ref{fig2}, column b). As more CO$_2$ is mixed in, it acquires more of a double peak structure, but the total width of the band remains the same. In pure H$_2$O ice the peak at  3696 cm$^{-1}$ (2.71 $\mu$m) attributed to the free/dangling OH stretch was not detected \citep{rowland91}. In contrast, Fig. \ref{fig2}, column a, shows that a free OH H$_2$O stretching band, centered around 3660 cm$^{-1}$ (2.73 $\mu$m), attributed to the stretches of small H$_2$O clusters, is clearly present in all ice mixtures with CO$_2$ and is comparable in intensity to the bulk stretching band at high concentrations of CO$_2$. While the intensity of the H$_2$O libration mode clearly drops as more CO$_2$ is mixed in, it is difficult to tell whether the profile of the libration band is affected by the presence of CO$_2$, because of the overlap with the CO$_2$ bending mode.

Of all the H$_2$O bands, only the free OH stretching band grows in strength as the amount of CO$_2$ in the ice increases. The band strengths of all other H$_2$O ice bands drop with increased concentration of CO$_2$, as illustrated in Fig. \ref{fig4}, where all band strengths have been scaled to that of the H$_2$O stretch in pure H$_2$O ice. The band strength of the bulk stretching band and the libration band are strongly dependent on the CO$_2$ concentration, while the intensity of the bending mode is less affected, whether the librational overtone is included or not, in calculating its integrated intensity. The relationship between the integrated peak intensities and CO$_2$ concentration is well described by linear models within this experimental domain and the data can be fitted with a typical squared correlation coefficient $R{^2}$ $=$ 0.98 (Table \ref{table4}). The correlations for the band strengths of the bending mode are considerably less, $R{^2}$ = 0.81 and 0.89 at 15 K and 45 K, respectively. This is due to the relatively small change in band strength with CO$_2$ concentration. 

In addition, the relative strengths of the various components of the bending modes were calculated from the previously fitted Gaussians. The integrated areas relative to the pure H$_2$O bending mode are listed in Table \ref{table3} and plotted in Fig. \ref{fig5} together with the area of the residual wing. The ratio between the 1661 and 1634 cm$^{-1}$ peaks is plotted and its significance in astrophysical applications is discussed in section 5.2. Table \ref{table3} also contains the total strength of the bending mode, excluding the wing. Due to the exclusion of the wing the drop in band strength with CO$_2$ concentration is smaller than in Fig. \ref{fig4}.

\begin{table*}
\caption{The defining parameters - peak position, FWHM (full width at half maximum) and peak height - of the fitted Gaussian components of the H$_2$O bending mode. In addition, the integrated area for each peak and their total sum are listed. The area of the libration overtone wing is also listed, but not included in the total area.}
\label{table3}
\centering
\begin{tabular}{l c c c c c }
\hline\hline
Ice mixture&Peak position &FWHM &Relative&Relative&Total integrated area\\
&[cm$^{-1}$ ($\mu$m)]&[cm$^{-1}$ ($\mu$m)]&peak height&integrated area&excl. wing\\
\hline
Pure H$_2$O&1661 (6.02)&130 (0.49)&1&1&1\\
&wing&&&0.54\\
\hline
H$_2$O:CO$_2$ 1:0.25&1609 (6.22)&15 (0.059)&0.056&0.0067&1.29\\
&1642 (6.09)&49 (0.19)&0.93&0.31&\\
&1661(6.02)&130 (0.49)&0.91&0.91&\\
&1685 (5.93)&35 (0.12)&0.22&0.055\\
&wing&&&0.60&\\
\hline
H$_2$O:CO$_2$ 1:0.5&1609 (6.22)&15 (0.059)&0.14&0.017&1.09\\
&1642 (6.09)&49 (0.19)&1.1&0.36&\\
&1661(6.02)&130 (0.49)&0.65&0.65&\\
&1685 (5.93)&35 (0.12)&0.22&0.055\\
&wing&&&0.32&\\
\hline
H$_2$O:CO$_2$ 1:1&1609 (6.22)&15 (0.059)&0.42&0.050&0.96\\
&1638 (6.11)&49 (0.19)&1.4&0.47&\\
&1661(6.02)&130 (0.49)&0.40&0.42&\\
&1685 (5.93)&35 (0.12)&0.22&0.055\\
&wing&&&0.07&\\
\hline
H$_2$O:CO$_2$ 1:2&1609 (6.22)&15 (0.059)&1.3& 0.15&0.99\\
&1636 (6.11)&49 (0.19)&1.6&0.54&\\
&1661(6.02)&130 (0.49)&0.27&0.27&\\
&1685 (5.93)&35 (0.12)&0.14&0.034\\
&wing&&&0.02&\\
\hline
H$_2$O:CO$_2$ 1:4&1609 (6.22)&15 (0.059)&2.8&0.34&0.94\\
&1634 (6.12)&49 (0.19)&1.7&0.57&\\
&1685 (5.93)&35 (0.12)&0.14&0.034\\
&wing&&&0.01&\\
\hline
\end{tabular}
\end{table*}

\begin{table*}
\caption{The linear fit coefficients for the H$_2$O band strengths as functions of CO$_2$ concentrations in \%, based on six experiments with 0 to 80\% CO$_2$. The last two rows show the linear fit to the ratio between the bulk stretching and the bending mode.}
\label{table4}
\centering
\begin{tabular}{l cccc}
\hline\hline
 &&\multicolumn{2}{c}{Linear Coefficients}\\
 Peak&Temperature&constant&linear coefficient&$R{^2}$\\
 &[K]&[10$^{-16}$ cm molecule$^{-1}$]&[10$^{-19}$ cm molecule$^{-1}$]&\\
\hline
H$_2$O libration&15&0.32$\pm$0.02&$-$3.2$\pm$0.4&0.99\\
&45&0.42$\pm$0.03&$-$2.7$\pm$0.6&0.92\\
H$_2$O bend&15&0.14$\pm$0.01&$-$0.5$\pm$0.2&0.81\\
&45&0.17$\pm$0.01&$-$0.6$\pm$0.1&0.89\\
H$_2$O bulk stretch&15&2.1$\pm$0.1&$-$22$\pm$2&0.99\\
&45&2.8$\pm$0.1&$-$21$\pm$2&0.98\\
H$_2$O free OH stretch&15&0&1.62$\pm$0.07&0.99\\
&45&0&1.40$\pm$0.05&0.99\\
\hline
&&constant &linear coefficient&$R{^2}$\\
\hline
H$_2$O bulk stretch / bend&15&16.7$\pm$0.2&$-$0.160$\pm$0.005&0.99\\
&45&17.0$\pm$0.3&$-$0.101$\pm$0.008&0.99\\
\hline
\end{tabular}
\end{table*}

\begin{table}
\begin{minipage}[t]{\columnwidth}
\caption{Ratios between the 1634 and 1661 cm$^{-1}$ components for all mixtures at 15, 45 and 75 K}
\label{table5}
\centering
\renewcommand{\footnoterule}{}
\begin{tabular}{l c c c}
\hline\hline
 &\multicolumn{3}{c}{Ratio at given temperature}\\
 Composition&15 K&45 K&75 K\\
\hline
pure H$_2$O&0&0&0\\
H$_2$O:CO$_2$ 1:0.25&0.34&0.12&0.034\\
H$_2$O:CO$_2$ 1:0.5&0.55&0.28&0.11\\
H$_2$O:CO$_2$ 1:1&1.1&0.49&0.11\\
H$_2$O:CO$_2$ 1:2&2.0&0.73&0.17\\
H$_2$O:CO$_2$ 1:4&$>$10&1.6&0.31\\
\hline
\end{tabular}
\end{minipage}
\end{table}

\subsection{Temperature dependence}

\begin{figure*}
\centering
\includegraphics[width=17cm]{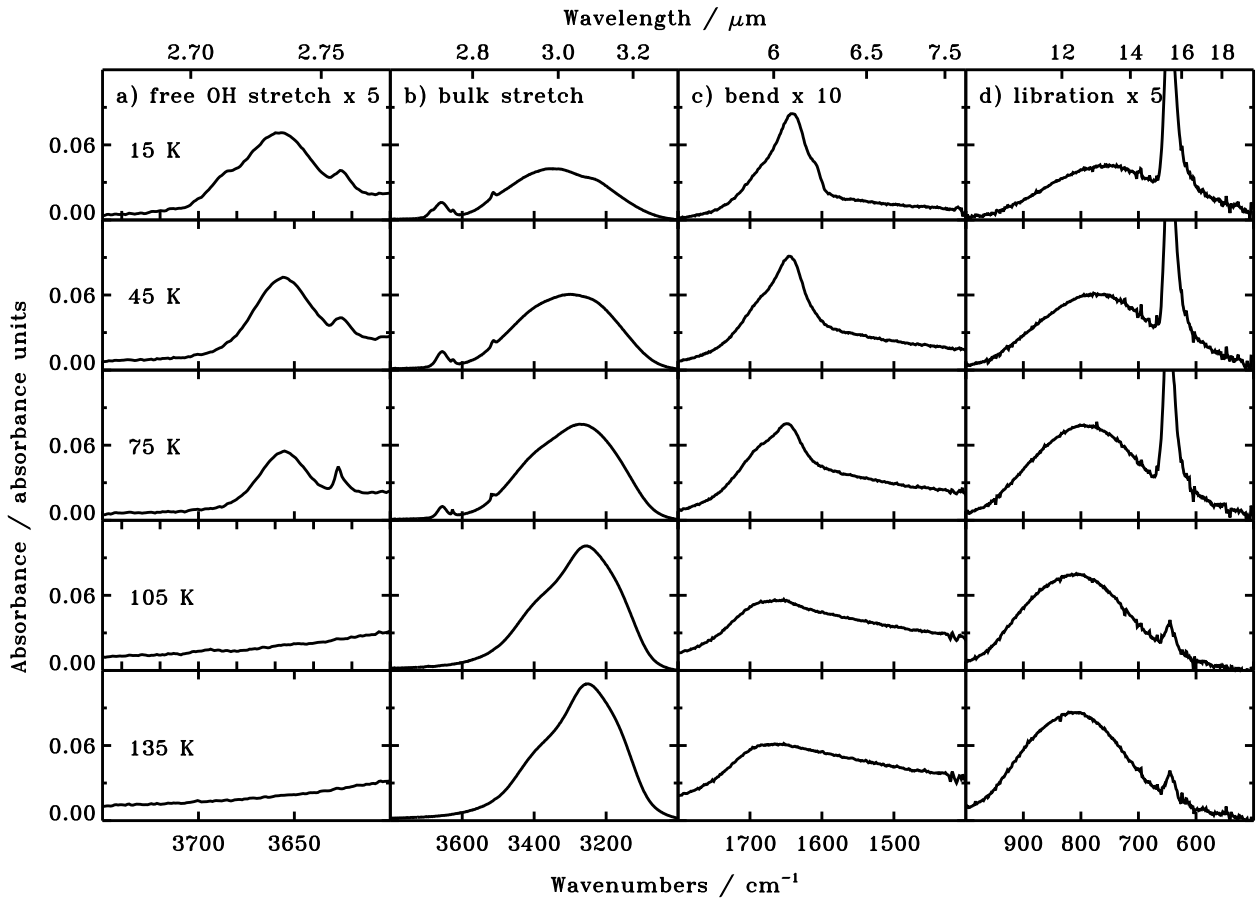}
\caption{Temperature dependence of the H$_2$O bands in a 1:0.5 H$_2$O:CO$_2$ mixture i.e. 67\% H$_2$O and 33\% CO$_2$. Spectra at 4000-400 cm$^{-1}$ of the four main H$_2$O peaks in a 1:0.5 ice mixture with CO$_2$ at temperatures between 15 and 135 K. Each column contains one of the H$_2$O vibrational modes and each row one of the temperatures. a) an expanded view of the H$_2$O free OH stretch, b) the H$_2$O bulk and free-OH stretch, c) the H$_2$O bend and  d) H$_2$O libration. The spectra of libration, bending and free OH stretch have been scaled, with a scaling factor indicated in the top row. }
\label{fig6}
\end{figure*}

\begin{figure}
\resizebox{\hsize}{!}{\includegraphics{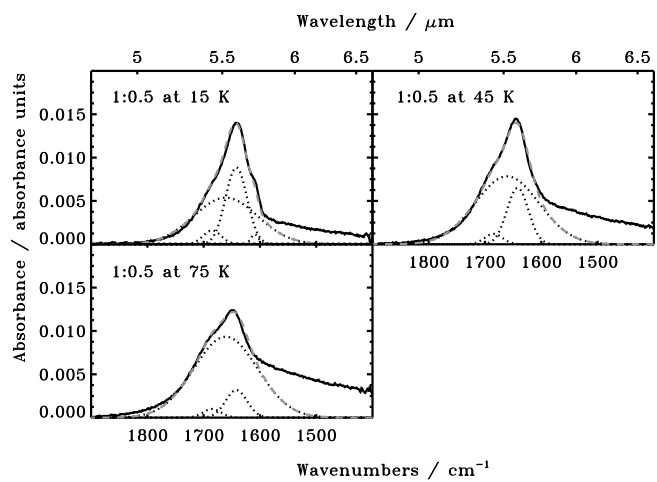}}
\caption{The components present in the bending mode in the H$_2$O:CO$_2$ 1:0.5 mixture at different temperatures. The same components that were derived in Fig. \ref{fig4} have been used to fit the bending mode at 15, 45 and 75 K. These components are always present regardless of temperature, but the ratios between the different components changes, especially the ratio between the thin 1634 cm$^{-1}$ and the pure H$_2$O 1661 cm$^{-1}$ components decreases with temperature.}
\label{fig7}
\end{figure}

\begin{figure}
\resizebox{\hsize}{!}{\includegraphics{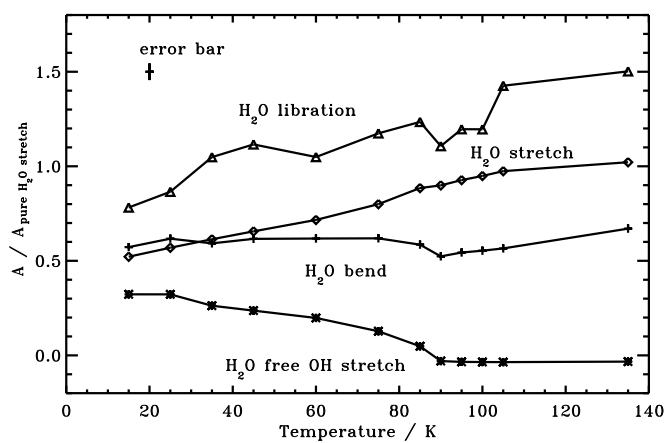}}
\caption{The band strengths of the four main H$_2$O peaks in an initial 1:0.5 ice mixture with CO$_2$ at temperatures between 15 and 135 K relative to the band strength of the H$_2$O bulk stretch in pure H$_2$O ice at 15 K. The band strengths of the libration, bending and free-OH stretching mode have been scaled by 10 to facilitate viewing. The average error bar for the relative band strengths is shown in the upper left corner.}
\label{fig8}
\end{figure}

\begin{figure}
\resizebox{\hsize}{!}{\includegraphics{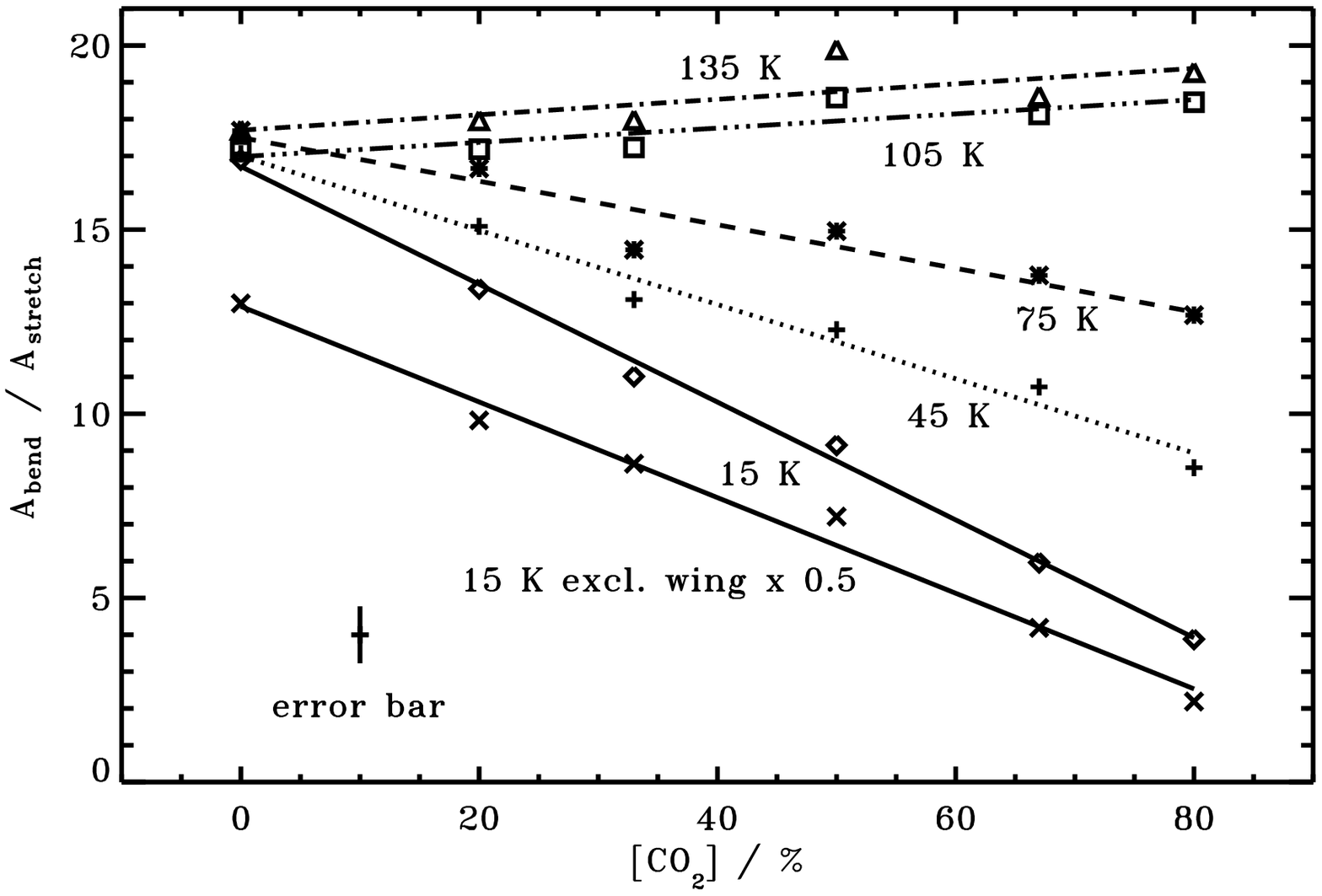}}
\caption{The ratio between the H$_2$O bulk stretching and bending mode band strengths (including the libration overtone wing) for different temperatures in H$_2$O:CO$_2$ ice mixtures with increasing initial amounts of CO$_2$. The CO$_2$ concentration quoted is that of the initial ice mixture, while above 90 K most of the CO$_2$ has desorbed, changing the composition substantially. In addition the ratio between the stretching band and the bending mode components (excluding the libration overtone wing) is shown for 15 K (multiplied by 0.5 to facilitate viewing).}
\label{fig9}
\end{figure}

The profiles of all H$_2$O bands in the H$_2$O:CO$_2$ mixtures change systematically with temperature to become more similar to the bands of pure H$_2$O ice as the temperature is increased (Fig. \ref{fig6}). This is especially apparent for the bending mode, where the narrow bands lose intensity and the broad band, associated with pure H$_2$O ice, gains intensity as the temperature rises. This is also seen in Fig. \ref{fig7} where the bending mode of the 1:0.5 mixture at 15, 45 and 75 K has been decomposed into the same components derived from the pure H$_2$O and 1:4 mixture at 15 K. As the temperature increases the shapes of the components remain the same, but the ratio between the thin components and the pure H$_2$O component decreases. The ratios for all mixtures at 15, 45 and 75 K are listed in Table \ref{table5}. 

The strengths of all H$_2$O bands increase with temperature, except for the free OH stretching band which decreases and disappears completely above 90 K. Figure \ref{fig8} shows that the bulk stretching band increases monotonically in band strength, while the bending and libration bands display jumps and local minima. These jumps are most pronounced around the CO$_2$ desorption temperature ($\sim$90 K). As in the case of concentration dependency, the band strengths of the stretching and libration bands are more affected by temperature than the band strength of the bending mode (including or excluding the libration overtone wing). Note that the profile of the bending mode is the most affected by both changes in temperature and CO$_2$ concentration, however.

Different bands are thus differently affected by an increasing temperature. In addition, the spectra of ice mixtures with higher CO$_2$ concentration are more affected by changes in temperature than those of ices with less CO$_2$; the influence of CO$_2$ concentration on the intensity ratio between the bulk stretching and the bending modes (including the libration overtone wing) is plotted in Fig. \ref{fig9} for different temperatures between 15 and 135 K. In this plot the CO$_2$ concentration on the horizontal axis is the initial one. Above 90 K most of the CO$_2$ has desorbed, which changes the mixture composition. In addition, the ratio between the stretching mode and the sum of the bending mode components (excluding the wing) at 15 K is plotted for comparison. The parameters of the linear fit to the 15 and 45 K measured points are included in Table \ref{table4}.

\subsection{Dependence on additional parameters: deposition temperature and ice thickness}

\begin{figure}
\resizebox{\hsize}{!}{\includegraphics{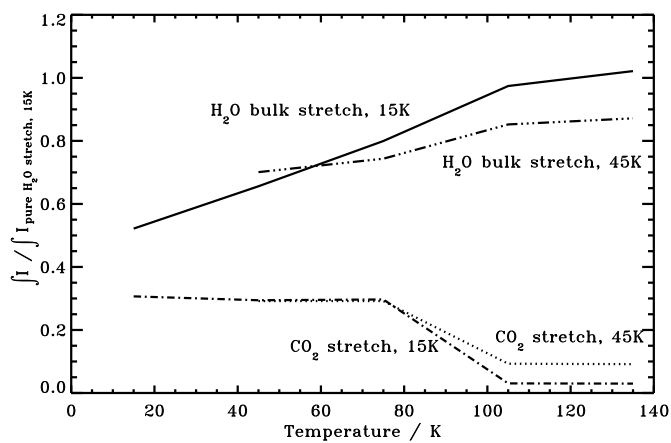}}
\caption{The band strengths of the H$_2$O stretching mode deposited at 15 and 45 K in the 1:0.5 ices, respectively, relative to the band strength of the H$_2$O stretch in pure H$_2$O ice. The CO$_2$ stretching peak  is plotted so the amount of CO$_2$ in the ice at different temperatures can be compared. The temperature labels given in the figure indicate the deposition temperature.}
\label{fig10}
\end{figure}

When the ice mixture (H$_2$O:CO$_2$ 1:0.5) was deposited at 45 K instead of 15 K the profiles and band strengths of the ice bands did not change compared to those found when heating the 15 K mixture to 45 K. The profiles remain similar also at higher temperatures. As the ices are heated, the increase in H$_2$O bulk stretching band strength is significantly smaller for the ice deposited at 45 K compared to 15 K. In addition, 2-3 times as much CO$_2$ is retained in the H$_2$O ice at temperatures above the CO$_2$ desorption temperature of 85-90 K (Fig. \ref{fig10}).

Two control experiments with approximately three times more and three times less total ice thickness were run to test for changes in peak profiles and relative peak band strengths with thickness (not shown here). The relative peak band strengths did not change significantly for these experiments (i.e. less than the previously estimated experimental uncertainty of 10\%) . Nevertheless, the profile showed some thickness dependence, with the narrow peak in the bending mode being more pronounced for the experiments with less ice coverage.

\section{Discussion}

\subsection{Ice structure}

The changes in the H$_2$O spectral features in H$_2$O:CO$_2$ ice mixtures, compared to the pure H$_2$O ice, demonstrate that the mixed-in CO$_2$ does affect the H$_2$O ice structure. The nature of the interaction between H$_2$O and CO$_2$ is not obvious. One possible scenario is that the CO$_2$ is spread out in the ice more or less uniformly and that the change in the H$_2$O peak intensities and profiles is due H$_2$O bonding with CO$_2$. A second option is that instead CO$_2$ forces the H$_2$O into small clusters, either in the gas-phase or upon arrival at the ice surface, since it is known from matrix isolated experiments and calculations that H$_2$O molecules form much stronger bonds with each other than with CO$_2$ \citep{tso85, danten05}. The spectral changes would then originate from a different type of H$_2$O-H$_2$O interaction rather than H$_2$O bonding with CO$_2$.

The large decrease in band strength of the H$_2$O bulk stretching mode that is observed when CO$_2$ is mixed into the H$_2$O ice indicates that CO$_2$ destroys the bulk hydrogen-bond network, since the band strength of the stretching mode is weaker in small clusters compared to that in larger clusters and bulk ice \citep{vanthiel57}. At the same time a new stretching peak appears at a higher frequency, i.e. more and more H$_2$O molecules are forced into a looser bound environment, which increases the intramolecular O-H bond strength, as the CO$_2$ concentration increases. The frequency of this band, $\sim$2.7 $\mu$m, and its distinctness from the bulk stretching band make it possible to assign it to free OH stretches \citep{rowland91}, which in general occur both at surfaces of ice and in clusters. This agrees with more and more H$_2$O molecules isolated in clusters, and hence separated from the H$_2$O ice hydrogen-bond network as CO$_2$ is added, but not with a uniform arrangement with CO$_2$. \citet{ehrenfreund96} observed a similar band in different ice mixtures with 10\% or less H$_2$O. They assigned it to overlapping peaks of H$_2$O monomers, dimers and small multimers consistent with our analysis. Similar conclusions were also drawn by \citet{vanthiel57} based on their cluster experiments.

Simultaneously with the appearance of the free OH stretching band, the original broad band in the bending region drops in intensity and two narrow peaks appear. The narrowest of the two dominates at high CO$_2$ concentration and is at the approximate position of the H$_2$O dimer in argon at 1611 cm$^{-1}$ \citep{ayers76}. However, the bending mode peaks of H$_2$O monomers, dimers and small clusters are not far apart and shift in position between different matrices \citep{tso85,vanthiel57}, thus rather than being due to only H$_2$O dimers it is more likely that this peak is produced by a mixture of monomer, dimer and small multimer peaks that are overlapping. The H$_2$O-CO$_2$ dimer has its main peak around 1598 cm$^{-1}$ \citep{tso85} and is not visible, further indicating that increasing the amount of CO$_2$ lead to a majority of the H$_2$O clustered with themselves rather than bonded with CO$_2$.

\citet{ehrenfreund96} observed a similar bending mode to ours in a H$_2$O:N$_2$:O$_2$ 1:5:5 mixture with a narrow peak at 1606 cm$^{-1}$ and a broader feature at 1630 cm$^{-1}$ assigned to H$_2$O monomers, dimers and small clusters and bulk H$_2$O ice, respectively. The fact that the same bending mode is acquired when H$_2$O is mixed with other small molecules corroborates our interpretation that the H$_2$O spectrum is dominated by the interaction between the H$_2$O molecules  regardless of mixture composition, as long as the mixed-in molecules cannot form hydrogen bonds, or more generally, cannot form bonds with H$_2$O of comparable strength to the H$_2$O-H$_2$O bond. Matrix-isolation experiments show that e.g. H$_2$O-CO, another astrophysically relevant combination, forms hydrogen bonds, while H$_2$O-CO$_2$ does not \citep{tso85}.

As the ice is heated the multimer peaks disappear quickly and all other peaks become more like those of pure H$_2$O. As the temperature increases the H$_2$O molecules can reorientate on their lattice points to a greater degree and form larger and larger hydrogen-bonded clusters and finally entire networks. This leads to a very small fraction of free OH bonds compared to those taking part in hydrogen bonds and hence only bulk vibrations show up in the spectra at high temperatures. Above 90 K the desorption of CO$_2$ further allows the hydrogen-bonds to reform. The temperature affects the H$_2$O molecules directly as well as indirectly through its effect on the mobility and final desorption of CO$_2$.

This scenario generally holds independent of whether the ice is deposited at 15 or 45 K. At both temperatures the deposited ice should be amorphous and porous, resulting in some CO$_2$ molecules becoming trapped in the H$_2$O ice as the ice is heated and the pores collapse. The H$_2$O peak profiles are similar at all temperatures, but it is clear that more CO$_2$ is trapped inside the H$_2$O ice when the ice is deposited at 45 K compared to 15 K. Figure \ref{fig10} shows that the higher fraction of CO$_2$ coincides with a lower intensity of the H$_2$O bulk stretching band, indicating that the trapped CO$_2$ makes it more difficult for the hydrogen bond network to reform. It is not clear at this stage why the ice deposited at 45 K is capable, when heated, of trapping more CO$_2$ within the matrix, but it is most likely a kinetic effect, related to the compactness of the ice. The different ice structure then affects the relative rates at which pores collapse versus CO$_2$ diffusion and desorption during the heating process \citep{collings03}.

All experimental results are hence consistent with a model in which the H$_2$O molecules are present in the ice either as bulk H$_2$O ice or as small H$_2$O clusters. A higher CO$_2$ concentration forces more H$_2$O molecules into the cluster state. Since H$_2$O bulk ice spectra are different from H$_2$O cluster spectra, adding CO$_2$ to H$_2$O ice will significantly change the shapes and band strengths of the different H$_2$O peaks compared to pure H$_2$O ice. It is expected that other small molecules that cannot form H-bonds (or only very weak ones), e.g. O$_2$ and N$_2$, will affect H$_2$O similarly to CO$_2$, as indicated by the matrix experiments of \citet{ehrenfreund96}. Our group is currently conducting a systematic series of experiment investigating the effects of N$_2$, O$_2$ and CO on H$_2$O spectral features in ice mixtures.

\begin{figure*} 
\centering 
\includegraphics[width=17cm]{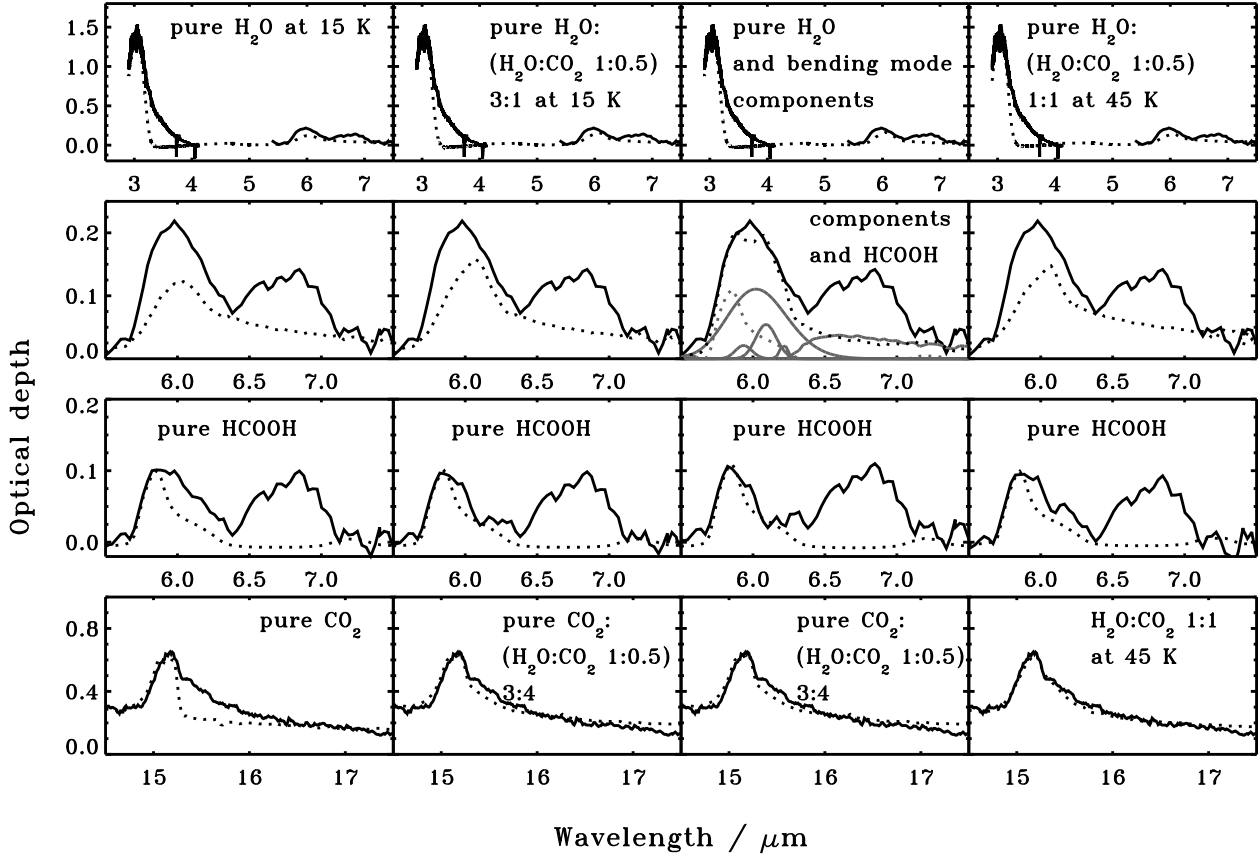}
\caption{Spectrum of the YSO B5 IRS1 (solid line) compared with laboratory spectra (dashed) of pure H$_2$O and CO$_2$ at 15 K (first column), a combination of pure H$_2$O and the H$_2$O:CO$_2$ 1:0.5 ice mixture at 15 K (second column) and 45 K (fourth column) and the derived bending mode components (third column). The silicate feature has been subtracted from the B5:IRS1 spectrum in all plots. In the upper row the spectrum has been fitted with respect to the 3 $\mu$m H$_2$O stretching band. The effect on the H$_2$O bending mode region is shown in the second row. In the third row the H$_2$O (mixture) spectra have been subtracted from the star spectrum and the residual is compared with a pure HCOOH spectrum at 15 K. The last row shows the fit to the CO$_2$ bending mode using the ice mixture derived from the H$_2$O modes. The CO$_2$ peak has been shifted in the laboratory spectra to account for the shift in position between C$^{16}$O$_2$ and C$^{18}$O$_2$ used in this study. In the fourth column the spectrum of pure HCOOH was fitted together with the bending mode components to the 6.0 $\mu$m band, to achieve an optimal fit while also being consistent with the band strength of the H$_2$O stretching mode. The individual components are plotted with a gray line.}
\label{fig11}
\end{figure*}

\begin{table*}
\begin{minipage}[t]{17cm}
\caption{Column densities derived from the different fits to the B5:IRS1 spectra}
\label{table6}
\centering
\renewcommand{\footnoterule}{}
\begin{tabular}{l c c c c c c}
\hline\hline
Composition&N(H$_{2}$O$_{\rm total}$)\footnote{from H$_2$O stretching mode}&N(H$_{2}$O$_{\rm pure}$) / &N(H$_{2}$O$_{\rm mixed}$) /&N(CO$_{2}$ $_{\rm total}$) /&N(CO$_{2}$ $_{\rm pure}$) /&N(CO$_{2}$ $_{\rm mixed}$) /\\
&(cm$^{-2}$)&N(H$_{2}$O$_{\rm total})$&N(H$_{2}$O$_{\rm total})$&N(H$_{2}$O$_{\rm total})$&N(H$_{2}$O$_{\rm total})$&N(H$_{2}$O$_{\rm total})$\\
\hline
pure H$_2$O at 15 K\footnote{first column in Fig. \ref{fig11}}&2.1$\times$10$^{18}$&1&0&0.32&0.32&0\\
H$_2$O:(H$_2$O:CO$_2$ 1:0.5)&2.3$\times$10$^{18}$&0.75&0.25&0.23&0.10&0.13\\
= 3:1 at 15 K\footnote{second column in Fig. \ref{fig11}, preferred fit at 15 K}&&&&&&\\
H$_2$O:CO$_2$ $\sim$90:10\footnote{third column in Fig. \ref{fig11}}&2.2$\times$10$^{18}$&0&1&-&-&0.08-0.14\\
assuming 15 K&&&&&&\\
H$_2$O:(H$_2$O:CO$_2$ 1:0.5)&1.7$\times$10$^{18}$&0.5&0.5&0.20&0&0.20\\
= 1:1 at 45 K\footnote{fourth column in Fig. \ref{fig11}, preferred fit at 45 K}&&&&&&\\
\hline
\end{tabular}
\end{minipage}
\end{table*}

\subsection{Astrophysical implications}

Recent Spitzer observations of the CO$_2$ bending mode at 15 $\mu$m towards background stars indicate that 85\% of the CO$_2$ column density is in a water-rich ice \citep{knez05}. To fit all H$_2$O and CO$_2$ features consistently, a combination of H$_2$O:CO$_2$ 10:1 and 1:1 at 10 K (with the ratios of the two mixtures varying between 1:0 and 1.3:1 for different sources) was used to fit the CO$_2$ profile \citep{knez05}.  It is hence not unlikely that some of the observed H$_2$O ice is in mixtures with close to equal concentrations of CO$_2$ and H$_2$O.

This study shows that CO$_2$ in H$_2$O:CO$_2$ ice mixtures affects the H$_2$O spectrum in three observable ways:
\begin{enumerate}
\item a free OH stretch peak appears around 2.73 $\mu$m,
\item the profile of the H$_2$O bending mode changes dramatically,
\item the H$_2$O stretching $/$ bending mode strengths decrease linearly with CO$_2$ concentration.
\end{enumerate}

When the free OH stretching region at 2.7 $\mu$m is observable the presence/absence of that band provides a stringent upper limit on the amount of CO$_2$ mixed into the H$_2$O ice. This region was covered by ISO for high mass YSOs and will be observed by Akari (ASTRO-F). It is not covered by Spitzer and cannot be observed from the ground. Hence it is not possible to use this region of the spectrum to constrain the CO$_2$ concentration in H$_2$O ice in most lines of sight.

Instead we will here use only spectral changes (2) and (3) to analyze the Spitzer and Keck spectra of the YSO B5:IRS1 after the silicate absorption has been removed \citep{boogert04}. We first compare the B5:IRS1 spectrum with laboratory spectra of pure H$_2$O, CO$_2$ ices at 15 K for reference. Then we deduce the maximum amount of H$_2$O that can be present in a H$_2$O:CO$_2$ 1:1 mixture without a change in peak position for the H$_2$O bending mode and then we present the best fit achieved by simultaneously comparing the H$_2$O stretching and bending regions and the CO$_2$ bending mode to laboratory spectra of pure H$_2$O, CO$_2$ and H$_2$O:CO$_2$ mixtures at 15 K. An equally good fit is achieved using laboratory spectra acquired at 45 K, which is also shown. In addition, we show that the average H$_2$O:CO$_2$ mixture in the line of sight can be deduced by fitting the 6.0 $\mu$m region with the bending mode components in Fig. \ref{fig3}, while keeping the total H$_2$O bending mode integrated intensity consistent with that of the H$_2$O stretching mode. For all cases we also attempt to fit the excess in the H$_2$O bending region with HCOOH. The presence of HCOOH is strongly indicated by absorption features at 7.25 and 8.2 $\mu$m (8.2 $\mu$m not shown). HCOOH has also been suggested before as the most likely candidate for the blue wing of the 6.0 $\mu$m band \citep{schutte96}.

In the analysis of B5:IRS1 we did not take into account the effect of grain shapes on the absorption profiles. The effect on the H$_2$O bending profile should be small, however, due to that the imaginary part of the optical constant (k) for the H$_2$O bending profile is small. When k is small it induces (through Kramers-Kronig relation) only small fluctuations of n around the central peak wave number so that particle scattering will not induce a significant change in band shape compared to transmission spectra. 

In Fig. \ref{fig11} the laboratory spectrum of pure H$_2$O is scaled to the 3.0 $\mu$m stretching band (upper row, first column). The fit in the bending mode region is shown in the second row and the spectrum after subtraction of the pure H$_2$O spectrum in the third row. A pure HCOOH spectrum was then scaled to the residual.  The lower row shows the fit of the CO$_2$ bending mode when only a pure CO$_2$ spectrum is used. It is found that the fit between the pure laboratory ice spectrum and observed spectrum is poor. We subsequently tested the maximum amount of H$_2$O that can be mixed in the 1:1 mixture with CO$_2$ without visibly changing the H$_2$O bending mode profile. If 6\% of the total H$_2$O column density is in a 1:1 mixture there is a clear change in peak position of the bending mode so 5\% was set as an upper limit.

In the second column a combination of pure H$_2$O and the H$_2$O:CO$_2$ 1:0.5 mixture at 15 K was used to fit the H$_2$O stretching and bending regions simultaneously. A good fit is also possible using the H$_2$O:CO$_2$ 1:1 and 1:0.25 mixture. The best fit is achieved using 3 parts pure H$_2$O ice and 1 part 1:0.5 mixture. The pure/mixture combination results in a smaller residual in the H$_2$O bending region when subtracted from the B5:IRS1 spectrum, compared to when only a pure H$_2$O spectrum is used. In addition, the fit with the spectrum of HCOOH to this residual is better. The same 1:0.5 ice mixture, together with pure CO$_2$, is also used to fit the CO$_2$ 15.2 $\mu$m bending mode. Approximately 50\% of the CO$_2$ is found to be in a H$_2$O rich ice for the best fit. A combination of pure and 1:0.25 mixture also results in a good fit for the CO$_2$ bending mode, but a combination of pure ice and the 1:1 mixture. Hence fitting the H$_2$O and CO$_2$ ice modes simultaneously effectively constrains both the abundances and different environments of both H$_2$O and CO$_2$ ice.

The third column shows the best fit to the 6.0 $\mu$m region using the H$_2$O bending mode components derived previously and pure HCOOH. This method of estimating the composition of water rich ice has the advantage that no mix and matching with different laboratory spectra is needed and the uncertainty is easier to estimate. The three components were varied independently, but the wing residual was scaled to the pure H$_2$O component. The ratio between the broad pure H$_2$O peak at 6.02 $\mu$m (1661 cm$^{-1}$) and the 6.09 $\mu$m (1642 cm$^{-1}$) peak was used to constrain the amount of CO$_2$ present by comparing this ratio to Fig. \ref{fig5}. The total H$_2$O bending mode integrated intensity was constrained by the ratio between the stretching and bending mode band strengths including the wing (Table \ref{table4}) and the ratio between the broad pure H$_2$O component and the narrow H$_2$O cluster bending component (Table \ref{table5}). The latter comparison constrains the amount of H$_2$O mixed with CO$_2$, which must be fed back into the ratio between the stretching and the total bending mode strength to calculate a correct column density. A good fit was achieved for average H$_2$O:CO$_2$ mixtures 92:8 to 86:14, which is consistent with the derived composition from the best fit in the second column. When higher resolution spectra of this region will be available, it should be possible to analyze all the different components of the H$_2$O bending mode as demonstrated in the laboratory spectrum (Fig. \ref{fig3}). This will further constrain the mixture composition from the ratio between the two narrow bending peaks apparent in all mixture spectra. In the current spectrum the ratio between the two narrow peaks only puts an upper limit  of 1:1 on the dominant H$_2$O:CO$_2$ mixture in the line of sight. The column densities derived from the fit of laboratory spectra and bending mode components are shown in Table \ref{table6}.

All the fits above were done with spectra at 15 K. If the observed ice is at a higher temperature, the same amount of CO$_2$ mixed into the ice will result in a H$_2$O bending mode profile that is less different from pure H$_2$O ice (Fig. \ref{fig6}). The best fit at 45 K has an average mixture of H$_2$O:CO$_2$ $\sim$ 1:0.23 (with 50\% of the H$_2$O ice in a 1:0.5 mixture and 50\% pure H$_2$O) and then no additional pure CO$_2$ is necessary to fit the CO$_2$ bending mode. A similar result is achieved when fitting the H$_2$O bending mode components and calculating the composition using the values of the component ratios for 45 K instead of 15 K in Table \ref{table5}. More generally the ratio of the components constrains a region in the ice mixture-temperature space (initial mixture if the temperature is above 90 K) rather than determining an exact mixture composition. The fit to B5:IRS1 resulted in a ratio of the 1634 to 1661 cm$^{-1}$ components of 0.15. According to Table \ref{table5} this means that the temperature:composition is bound by 15 K:11$\pm$3\%, 45 K:23$\pm$3\% and 75 K:39$\pm$5\% CO$_2$ relative to the H$_2$O ice abundance. Using the components to fit the bending region rather than spectra is advantageous if the temperature is unknown, since the ratio between the components sets bounds on the temperature and composition simultaneously and fully reveals the uncertainties involved. In most sources it is possible to independently constrain the temperature, however. For B5:IRS1 \citet{boogert04} have concluded that the bulk of the ice must be colder than 50K. For these temperatures the ratio of the bending mode components results in an average H$_2$O:CO$_2$ mixture of 11 to 23\% CO$_2$ with respect to the H$_2$O ice. This can be further refined by more detailed radiative transfer modeling of the dust temperature in the source by fitting the spectral energy distribution. In contrast the total abundances of H$_2$O and CO$_2$ are only weakly dependent on temperature and can be determined within 15\% uncertainty.

Note that the strengths of the stretching and libration bands are affected by CO$_2$, which needs to be taken into account when deducing the H$_2$O column density from the optical depths of the H$_2$O bands if a considerable amount of the H$_2$O is in mixtures with CO$_2$. In the example shown here the drop in band strength of the stretching band is only $\sim$5\% compared to the pure H$_2$O ice (Table \ref{table6}). In the case where $\sim$50\% of the H$_2$O ice is in a 1:1 mixture with CO$_2$, as has been deduced for some background stars by \citet{knez05}, the band strength of the H$_2$O stretching mode would drop by as much as $\sim$25\% compared to pure H$_2$O ice and the inferred H$_2$O column density would be $\sim$33\% higher. 

Whether the results presented here can explain the ratio between the H$_2$O stretching and bending modes in all astrophysical sources has to be determined for each source individually because of the drastic change in the bending mode profile that accompanies the drop in ratio as the concentration of CO$_2$ is increased. In the case of B5:IRS1 the H$_2$O and CO$_2$ band shapes and intensities are well described by 50\% of the CO$_2$ and 25\% of the H$_2$O ice in a H$_2$O:CO$_2$ 1:0.5 ice mixture at 15 K and by 100\% of the CO$_2$ and 50\% of the H$_2$O ice in a H$_2$O:CO$_2$ 1:0.5 ice mixture at 45 K. These ratios in turn can provide interesting constraints on the H$_2$O and CO$_2$ formation routes.

\section{Conclusions}

\begin{enumerate}
\item The H$_2$O band profiles and strengths in astrophysically plausible mixtures with CO$_2$ are significantly different compared to pure H$_2$O ice. The changes in H$_2$O band profiles with CO$_2$ are greatest for the H$_2$O bending mode, while the stretching band displays more subtle changes.
\item The band strengths of all major H$_2$O bands depend linearly on CO$_2$ in the investigated domain of 20\% to 80\% of CO$_2$ in the ice mixture. The band strength ratio between the H$_2$O bulk stretching and bending modes is also linearly dependent on CO$_2$ concentration in astrophysically relevant H$_2$O:CO$_2$ mixtures such that a H$_2$O:CO$_2$ 1:1 mixture has a ratio of half that of pure H$_2$O.
\item The H$_2$O bending mode in H$_2$O:CO$_2$ mixtures can be separated into three components due to pure H$_2$O ice, large H$_2$O and small H$_2$O clusters, respectively. In addition, a librational overtone overlaps with the bending mode.
\item The H$_2$O bending profile is visibly affected by as little as 6\% H$_2$O in a H$_2$O:CO$_2$ 1:1 mixture. The composition of the H$_2$O rich ice is hence crucial to identify other species contributing to the 6 $\mu$m band.
\item A free OH stretching band appears with CO$_2$ concentration which can be used to put strict upper limits on CO$_2$ mixed with H$_2$O whenever the 2.7 $\mu$m region is observed.
\item If the ice temperature can be estimated independently, the total amount of CO$_2$ and H$_2$O as well as the amount of CO$_2$ mixed with the H$_2$O ice can be constrained by fitting the appropriate laboratory mixture spectra consistently to the H$_2$O and CO$_2$ bands.
\item The average H$_2$O:CO$_2$ ice mixture in any line of sight can be efficiently deduced by fitting the H$_2$O bending mode with components corresponding to pure and CO$_2$ rich ice. Especially if the ice temperature is uncertain, fitting the H$_2$O bending mode components to the bending mode region together with HCOOH, will effectively constrain the composition and temperature simultaneously.
\item The deduction of mixture composition is substantially dependent on the assumed dust temperature. The determination of total column densities of H$_2$O and CO$_2$ are only weakly dependent on the assumed dust temperature, however.
\end{enumerate}

\begin{acknowledgements}

We thank Claudia Knez for initiating the discussion that prompted this study. We also thank the referee, Bernard Schmitt, for helpful comments on the manuscript. Funding was provided
by NOVA, the Netherlands Research School for Astronomy, a grant from the European Early Stage Training Network ('EARA' MEST-CT-2004-504604) and a NWO Spinoza grant.

\end{acknowledgements}

\end{document}